\documentclass[intlimits,twoside,a4paper]{article}

\usepackage[cp1251]{inputenc}

\usepackage{mathrsfs}
\usepackage[eqsecnum]{cmpj3}

\issue{2024}{27}{4}{43701}
\doinumber{10.5488/CMP.27.43701}

\title[Plasmonic resonances in the chain of spheroidal metallic nanoparticles]%
{Plasmonic resonances in the chain of spheroidal metallic nanoparticles on the dielectric substrate}
\author[M. S. Maniuk, A. V. Korotun, V. I. Reva, I. M. Titov]{M. S. Maniuk\orcid{0009-0006-4195-5577}\refaddr{label1},
        A. V. Korotun\orcid{0000-0003-4165-2788}\refaddr{label1,label2}\thanks{Corresponding author:\email{andko@zp.edu.ua}.},
        V. I. Reva\orcid{0000-0002-3265-1735}\refaddr{label1},
        I. M. Titov\orcid{0000-0000-0000-0000}\refaddr{label1}}
\addresses{
\addr{label1} National University ``Zaporizhzhia Politechnic'', 64 University St., Zaporizhzhya 69063, Ukraine
\addr{label2} G. V. Kurdyumov Institute for Metal Physics of the NAS of Ukraine, 36 Academician Vernadsky Blvd., Kyiv 03142, Ukraine
}
\Keywords{prolate metallic spheroids, chains of nanoparticles, dielectric substrate, chain polarizability tensor, plasmonic resonance, chain sums}

\date{Received February 4, 2024, in final form July 9, 2024}

\begin{document}

\maketitle

\begin{abstract}
The optical and plasmonic properties of the chains of prolate metallic spheroids of the dielectric substrate are studied in the work using the local field approximation. The case when spheroids are arranged in such a way that their major semi-axis belongs to the substrate plane is considered. The relations for the transverse component of the chain polarizability tensor and the frequency of the transverse chain resonance are obtained. The comparative analysis of spectral shifts of the maxima of the imaginary part of the transverse chain polarizability and the polarizability of isolated prolate spheroid is performed. The influence of the variation of the material of particles in the chain and the dielectric environment on the location of the maxima of the imaginary part of the transverse polarizability is studied. The limits of applicability of the theory proposed in the paper are established.
%
%
\printkeywords
%
\end{abstract}

\section{Introduction}


The assemblies of metallic nanoparticles make it possible to manipulate the light in compact photonic components using localized surface plasmonic resonances (SPR) \cite{B1}.
The capability of metallic nanoparticles to limit and increase the light intensity makes it possible to use them as the elements for the development of sensors \cite{B2}, nanolasers \cite{B3} and  interfaces for the extended Raman spectroscopy \cite{B4}. In particular, experimental \cite{B5}, analytical \cite{B6} and numerical \cite{B7} studies have considered nanoparticle chains for the subwavelength laterally confined light guides.

The waveguiding properties of the chains of closely arranged metallic nanoparticles (plasmonic chains) have been widely discussed in the literature during the last decade \cite{B8,B9,B10,B11}. Such systems can be synthesized, for example, using chemical self-assembly method \cite{B12,B13} or by various lithography techniques~\cite{B14,B15}. The nanoparticle chains can be useful for transmitting modulated optical signals with a high degree of spatial confinement. Localization of the electromagnetic fields is an important factor in the applications connected with the development of the nanoscale optical elements. For example, subwavelength spatial localization of the optical excitations in plasmonic chains can be expected to minimize parasitic interactions between different elements of the optical circuit. Alternative nanoscale waveguide constructions, including nanowires \cite{B16} and nanochannels on the plane surface \cite{B16,B17}, as well as stripes, ridges and other similar structures~\cite{B18}, have also been considered in the literature. Currently, it is hard to say which construction can be considered the most promising. However, plasmonic chains are characterized by the exceptional capability of tuning the physical properties.

Optical signals propagate in the plasmonic chains in the form of surface plasmon-polaritons (SPP), which are the collective excitations of conduction electrons and electromagnetic field. SPP can be characterized by the dispersion law, according to which one can determine group and phase velocities. The dispersion relations for relatively long plasmonic chains, which consist of $N \approx {10^3}$ particles, have been widely studied for both the spherical particles \cite{B8, B19, B20}, and for the particles of nonspherical forms~\cite{B9}. In particular, it has been shown that group velocities of SPP in the chains of spherical particles are much less than the light velocity. Hence, the bandwidth of such waveguides is very limited. This problem can be solved using nonspherical particles \cite{B9, B21}. In plasmonic chains of spheroidal particles with sufficiently small aspect ratio, group velocities can be of the order of the speed of light. Thus, the spectral interval, where the dispersion law is close to linear can be large enough to allow the wave packets to propagate \cite{B9}. However, ohmic losses strongly effect the SPP in the long chains. There are constructions in which the transmission of wave packages along the long chains is possible almost without amplitude attenuation \cite{B22, B23}, but these constructions are energetically consumable on the first particle in the chain, which can cause a significant heating.

For this and other reasons, relatively short chains are of interest. In plasmonic chains, which consist of $N \leqslant 20$ particles, ohmic losses are negligible, which is an important property for  practical applications. In addition, short chains may be required for miniaturization purposes. The dispersion patterns in relatively short plasmonic chains of spherical particles have been considered in the literature~\cite{B24}. However, short plasmonic chains have disadvantages. In particular, there is a parasitic effect of SPP reflections from the ends of the chain. The propagation of SPP wave packets in  short chains, especially when the chains consist of nonspherical particles, has received little attention so far.

Electromagnetic energy transfer along the chains of metallic nanoparticles is based on the near-field electrodynamic interaction between metallic particles, which creates bound dipole plasmonic modes~\cite{B25}. This type of coupling is analogous to the process of the resonant energy transfer, which is observed in the systems which contain closely located optically excited atoms, molecules or semiconductor nanocrystals~\cite{B26}.

It should be pointed out that plasmons in the chains of metallic nanoparticles of different shapes (spherical particles, oblate spheroids, resonators) on a dielectric substrate were studied in the works~\cite{B7, B22, B23, B27, B28}. However, the issue of excitation of plasmonic resonances in the chain of prolate spheroids has not been investigated. In this connection, the study of the optical and plasmonic properties of such chains is an urgent task.

\section{Basic relations}

\begin{figure}[htb]
\centerline{\includegraphics[width=0.65\textwidth]{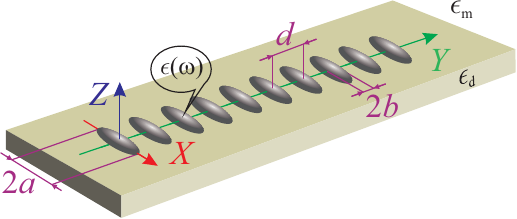}}
\caption{(Colour online) Geometry of the problem.} 
\label{fig1}
\end{figure}

Let us consider the chain of prolate metallic nanospheroids located on a dielectric substrate in such a way that their major semi-axis lies in the plane of the substrate (figure~\ref{fig1}). We  use the local field approximation \cite{B29}, which takes into account the interaction of nanospheroids among themselves and with image dipoles. In the above approximation, under the condition of normal incidence of light on the substrate, the transverse component of the effective (chain) polarizability tensor is determined formally by the same relation as in the case of the two-dimensional lattice of nanospheroids on a dielectric substrate

\begin{equation}
\alpha _{\rm{chain}}^ \bot  = \frac{\alpha^{\bot }\left( \omega  \right)}{{1 - \frac{\alpha ^{ \bot }\left( \omega  \right)}{{\epsilon_{\rm{m}}}{d^3}}}\left( {{\cal S}_ \bot ^d + \frac{{{\epsilon_{\rm{d}}} - {\epsilon_{\rm{m}}}}}{{{\epsilon_{\rm{d}}} + {\epsilon_{\rm{m}}}}}{\cal S}_ \bot ^i} \right)},
\label{eq1}
\end{equation}
where $d$ is the period of the chain (the distance between the centers of spheroids); $\epsilon_{\rm{d}}$ and $\epsilon_{\rm{m}}$ are  dielectric permittivities of the material of the substrate and environment, correspondingly; ${\cal S}_ \bot ^d$ and ${\cal S}_ \bot ^i$ are the chain sums, which are determined by the contributions of the other particles of the chain and image dipoles; transverse component of  polarizability tensor of prolate spheroids

\begin{equation}
{\alpha^\bot }\left( \omega  \right) = V\frac{{\epsilon^ \bot }\left( \omega  \right) - {\epsilon_{\rm{m}}}}{{{\epsilon_{\rm{m}}} + {{\cal L}_ \bot }\left( {{\epsilon^ \bot }\left( \omega  \right) - {\epsilon_{\rm{m}}}} \right)}}.
\label{eq2}
\end{equation}

In formula \eqref{eq2}: $V = {{4\piup a{b^2}} \mathord{\left/
 {\vphantom {{4\pi a{b^2}} 3}} \right.
 \kern-\nulldelimiterspace} 3}$ is the volume of the spheroidal nanoparticle; depolarization factors are as follows

\begin{equation}
{{\cal L}_ \bot } = \frac{1}{2}\left( {1 - {{\cal L}_\parallel }} \right);
\,\,\,\,
{{\cal L}_\parallel } = \frac{{1 - e_p^2}}{{2e_p^3}}\left( {\ln \frac{{1 + {e_p}}}{{1 - {e_p}}} - 2{e_p}} \right),
\label{eq3}
\end{equation}
where the eccentricity

\begin{equation}
e_{p} = \sqrt {1 - \frac{b^{2}}{a^{2}}},
\label{eq4}
\end{equation}
$a$ and $b$ are the major and minor semi-axes of spheroid;

\begin{equation}
{\epsilon^ \bot }\left( \omega  \right) = {\epsilon^\infty } - \frac{{\omega _p^2}}{{\omega \left( {\omega  + {\mathop{\rm i}\nolimits} \gamma _{{\rm{eff}}}^ \bot } \right)}}
\label{eq5}
\end{equation}
is the transverse component of the dielectric tensor of spheroid material in Drude model, ${\omega _p}$ is the plasma frequency; ${\epsilon^\infty }$ is the contribution of interband transitions into the dielectric permittivity, $\gamma _{\rm{eff}}^ \bot $ is the transverse component of the effective relaxation rate of electrons

\begin{equation}
\gamma _{\rm{eff}}^ \bot  = {\gamma _{{\rm{bulk}}}} + \gamma _{{\rm{surf}}}^ \bot  + \gamma _{{\rm{rad}}}^ \bot,
\label{eq6}
\end{equation}
where $\gamma _{\rm{bulk}} = {\mathop{\rm const}\nolimits} $ is the bulk relaxation rate, while the surface relaxation rate and the radiation attenuation rate are determined by the relations \cite{B30,B31}:

\begin{equation}
\gamma _{\rm{surf}}^ \bot  = \frac{9}{{16}}\frac{{{{\cal L}_ \bot }}}{{\epsilon_{\rm{m}} + {{\cal L}_ \bot }\left( {1 - {\epsilon_{\rm{m}}}} \right)}}\frac{{{v_{\rm{F}}}}}{{2b}}{\left( {\frac{{{\omega _p}}}{\omega }} \right)^2}{\mathscr{F}_ \bot }\left( {e_{p}} \right);
\label{eq7}
\end{equation}

\begin{equation}
\gamma _{{\rm{rad}}}^ \bot  = \frac{V}{{8\piup }}\frac{{{{\cal L}_ \bot }}}{{\sqrt {{\epsilon_{\rm{m}}}\left( {{\epsilon^\infty } + \frac{{1 - {{\cal L}_ \bot }}}{{{{\cal L}_ \bot }}}{\epsilon_{\rm{m}}}} \right)} }}{\left( {\frac{{{\omega _p}}}{c}} \right)^3}{\left( {\frac{{{\omega _p}}}{\omega }} \right)^2}\frac{{{v_{\rm{F}}}}}{{2b}}{\mathscr{F}_ \bot }\left( {{e_p}} \right).
\label{eq8}
\end{equation}
Here $c$ is the light velocity; ${v_{\rm{F}}}$ is the Fermi electron velocity, and the size-dependent function has the form

\begin{equation}
\mathscr{F}_ {\bot }\left( {{e_p}} \right) = \frac{1}{{e_p^3}}\left\{ {{e_p}\sqrt {1 - e_p^2} \left( {\frac{1}{2} + e_p^2} \right) + 2\left( {e_p^2 - \frac{1}{4}} \right)\left( {\frac{\piup }{2} - \arcsin \sqrt {1 - e_p^2} } \right)} \right\}.
\label{eq8}
\end{equation}

Let us determine the chain sums ${\cal S}_ \bot ^d$ and ${\cal S}_ \bot ^i$ for the chain of prolate spheroids. It is most convenient to do this  using the expressions for two-dimensional lattice sums and making the transition to the one-dimensional case. In this case, the expressions for the sums have the form

\begin{equation}
{\cal S}_ \bot ^d = 2\sum\limits_{n =  - \infty }^\infty  {\frac{1}{{{n^3}}}}  = 4\zeta \left( 3 \right);
\label{eq9}
\end{equation}
\begin{equation}
{\cal S}_ \bot ^i = \sum\limits_{n =  - \infty }^\infty  {\frac{{ - 2{n^2} + {{\left( {{{2{z_0}} \mathord{\left/
 {\vphantom {{2{z_0}} d}} \right.
 \kern-\nulldelimiterspace} d}} \right)}^2}}}{{{{\left[ {{n^2} + {{\left( {{{2{z_0}} \mathord{\left/
 {\vphantom {{2{z_0}} d}} \right.
 \kern-\nulldelimiterspace} d}} \right)}^2}} \right]}^{{5 \mathord{\left/
 {\vphantom {5 2}} \right.
 \kern-\nulldelimiterspace} 2}}}}}}  = \frac{{\piup d}}{{{z_0}}}\sum\limits_{m = 1}^\infty  {m\left\{ {{K_1}\left( {4\piup \frac{{{z_0}m}}{d}} \right) - \frac{{\piup d}}{{{z_0}}}{m^3}{K_2}\left( {4\piup \frac{{{z_0}}}{{{a_l}}}m} \right)} \right\}},
 \label{eq10}
\end{equation}
where $n$ and $m$ are the numbers numbering spheroidal nanoparticles in the chain; $\zeta \left( z \right)$ is the Riemann zeta function ($\zeta \left( 3 \right) \approx 1.202$ is the Apery constant); ${K_\nu }\left( x \right)$ is the McDonald function of the order $\nu $; ${z_0}$ is the distance between the substrate surface and the layer, which contain image dipoles.

The size dependencies for the frequency of the transverse surface chain resonance under the absence of attenuation (when $\gamma _{{\rm{eff}}}^ \bot  \to 0$) can be found from the condition of the equality to zero of the denominator of the expression \eqref{eq1}:

\begin{equation}
1 - \frac{{{\alpha ^ \bot }\left( \omega  \right)}}{{{\epsilon_{\rm{m}}}{d^3}}}\left( {{\cal S}_ \bot ^d + \frac{{{\epsilon_{\rm{d}}} - {\epsilon_{\rm{m}}}}}{{{\epsilon_{\rm{d}}} + {\epsilon_{\rm{m}}}}}{\cal S}_ \bot ^i} \right) = 0.
\label{eq11}
\end{equation}

Substituting the relations \eqref{eq2} and \eqref{eq3} into \eqref{eq11}, we obtain the size dependence

\begin{equation}
\omega _{{\rm{sp}}}^{{\rm{chain}}{\rm{,}}\, \bot } = \frac{{{\omega _p}}}{{\sqrt {{\epsilon^\infty } + \frac{{1 - {{\tilde {\cal L}}_ \bot }}}{{{{\tilde {\cal L}}_ \bot }}}{\epsilon_{\rm{m}}}} }},
\label{eq12}
\end{equation}
where renormalized depolarization factor is as follows:

\begin{equation}
{\tilde {\cal L}_ \bot } = {{\cal L}_ \bot } - \frac{V}{{{\epsilon_{\rm{m}}}{d^3}}}\left( {{\cal S}_ \bot ^d + \frac{{\epsilon_{\rm{d}} - \epsilon_{\rm{m}}}}{{{\epsilon_{\rm{d}}} + {\epsilon_{\rm{m}}}}}{\cal S}_ \bot ^i} \right).
\label{eq13}
\end{equation}

In what follows, relations \eqref{eq1} and \eqref{eq12} will be used to obtain numerical results, taking into account formulae \eqref{eq2}--\eqref{eq10} and \eqref{eq13}.

\section{Calculation results and their discussion}

Calculations for frequency dependencies of the transverse component of the polarizability tensor and the size dependence of the chain resonance frequency have been performed for prolate spheroids (and the chains of them) of different sizes and different metals. The parameters of metals and dielectrics are given in tables~\ref{tbl-smp1},~\ref{tbl-smp2}.

\begin{table}[htb]
\caption{Parameters of metals (see, for example, \cite{B30,B32} and the references to them).}
\label{tbl-smp1}
\vspace{1ex}
\begin{center}
\renewcommand{\arraystretch}{0}
\begin{tabular}{cccccc}
 \hline%
 \strut
 & \multicolumn{4}{c}{Metal}\\%
\cline{2-6}\raisebox{2ex}[0cm][0cm]{Parameters} \strut &Cu&Au&Ag&Pt&Pd\\
 \hline%
  \strut
${r_s}/{a_0}$&2.11&3.01&3.02&3.27&4.00\\%
${m^*}/{m_e}$&1.49&0.99&0.96&0.54&0.37\\%
${\epsilon^\infty }$&12.03&9.84&3.7&4.42&2.52\\%
${\gamma _{{\rm{bulk}}}},\,\,{10^{14}}\,\,{{\rm{s}}^{ - 1}}$&0.37&0.35&0.25&1.05&13.90\\%
\hline
\end{tabular}
\renewcommand{\arraystretch}{1}
\end{center}
\end{table}

\begin{table}[htb]
\caption{Dielectric permittivities of matrices (see, for example, \cite{B33} and the references to them).}
\label{tbl-smp2}
\vspace{1ex}
\begin{center}
\renewcommand{\arraystretch}{0}
\begin{tabular}{cccccc}
 \hline%
  \strut
{Substance}&Air&${\rm{Ca}}{{\rm{F}}_2}$&Teflon&${\rm{A}}{{\rm{l}}_2}{{\rm{O}}_3}$&${\rm{Ti}}{{\rm{O}}_2}$\\
 \hline%
  \strut
$\epsilon_{\rm{m}}$&1.0&1.54&2.3&3.13&4.0\\%
\hline
\end{tabular}
\renewcommand{\arraystretch}{1}
\end{center}
\end{table}

\begin{figure}[!t]
\centerline{\includegraphics[width=0.85\textwidth]{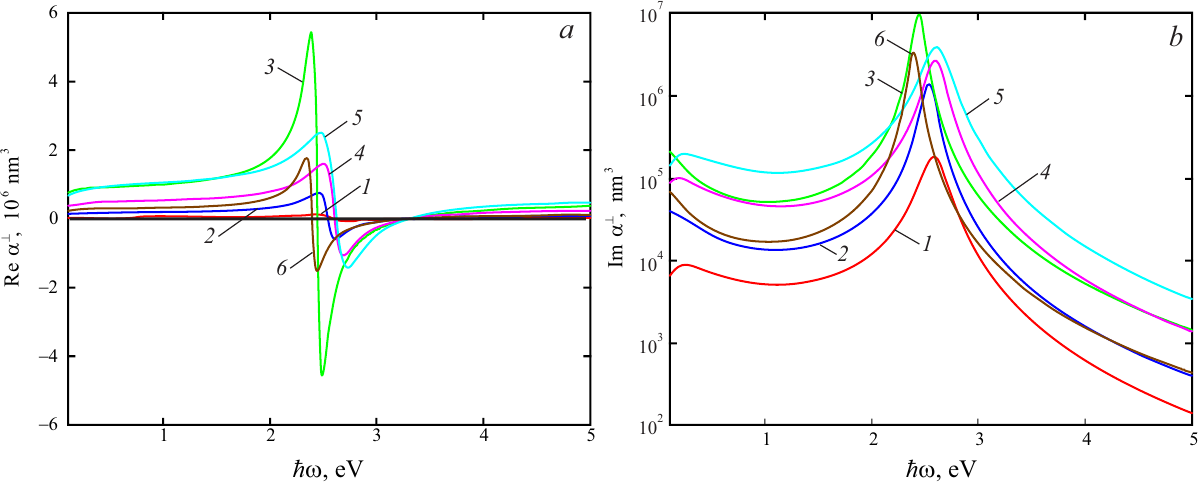}}
\caption{(Colour online) The frequency dependencies for the real (\emph{a}) and imaginary (\emph{b}) parts of the transverse component of the polarizability tensor of isolated prolate spheroids Au in Teflon: 1 --- $b = 10\,\,{\rm{nm}}$, $a = 50\,\,{\rm{nm}}$; 2 --- $b = 20\,\,{\rm{nm}}$, $a = 50\,\,{\rm{nm}}$; 3 --- $b = 40\,\,{\rm{nm}}$, $a = 50\,\,{\rm{nm}}$; 4 --- $b = 20\,\,{\rm{nm}}$, $a = 150\,\,{\rm{nm}}$; 5 --- $b = 20\,\,{\rm{nm}}$, $a = 300\,\,{\rm{nm}}$; 6 --- $a = b = 50\,\,{\rm{nm}}$ (sphere).}
\label{fig2}
\end{figure}

The frequency dependencies for the real and imaginary parts of the transverse component of the polarizability tensor of prolate spheroids Au are shown in figure~\ref{fig2}. It should be pointed out that 
$\Re \alpha^\bot $ is an alternating function of frequency, while 
$\Im \alpha^\bot >0$ in the entire frequency range which is under the consideration. The maxima of the imaginary part experience a ``red'' shift (curves in the sequence $1 \to 2 \to 3$) under a increase in the value of minor semi-axis of spheroid (under the constant value of major semi-axis). At the same time, the ``blue'' shift of $\max \left\{ {{\Im \alpha^\bot}} \right\}$ (curves in the sequence $2 \to 4 \to 5$) takes place under an increase in the value of the major semi-axis (under the constant value of minor semi-axis). This fact indicates that the frequencies of the transverse SPR (which correspond to the maximum of the imaginary part of the transverse polarizability component) of spheroids decrease as the shape of elongated spheroid approaches the spherical one. The limiting case of the sphere with $R=50$~nm  is represented by the curves~6. Let us point out that SPR frequency in this case is less than SPR frequency in the case of prolate spheroids.

The mentioned fact also follows from the estimation of the difference between SPR frequencies for spheroid and sphere under small eccentricities (${e_p} \to 0$). Since
\begin{equation}
\omega _{{\rm{sp}}}^{{\rm{sphere}}} = \frac{{{\omega _p}}}{{\sqrt {{\epsilon^\infty } + 2{\epsilon_{\rm{m}}}} }},
\,\,\,\,
\omega _{{\rm{sp}}}^{ \bot ,\,\,{\rm{prolate\,\,spheroid}}} = \frac{{{\omega _p}}}{{\sqrt {{\epsilon^\infty } + \frac{{1 - {{\cal L}_ \bot }}}{{{{\cal L}_ \bot }}}{\epsilon_{\rm{m}}}} }}
\label{eq14}
\end{equation}
then, taking into account expression \eqref{eq3}, we have

\begin{equation}
\frac{{1 - {{\cal L}_ \bot }}}{{{{\cal L}_ \bot }}} = \frac{{1 + {{\cal L}_\parallel }}}{{1 - {{\cal L}_\parallel }}},
\end{equation}
where

\begin{equation}
{{\cal L}_\parallel } = \frac{1}{3} + \frac{1}{5}e_p^2.
\label{eq15}
\end{equation}

Then,

\begin{equation}
\Delta {\omega _{sp}} = \omega _{{\rm{sp}}}^{ \bot ,\,\,{\rm{prolate\,\,spheroid}}} - \omega _{{\rm{sp}}}^{{\rm{sphere}}} = \frac{{{\omega _p}}}{{\sqrt {{\epsilon^\infty } + 2{\epsilon_{\rm{m}}}} }}\left[ {\sqrt {\frac{{{\epsilon^\infty } + 2{\epsilon_{\rm{m}}}}}{{{\epsilon^\infty } + \frac{{1 + {{\cal L}_\parallel }}}{{1 - {{\cal L}_\parallel }}}{\epsilon_{\rm{m}}}}}}  - 1} \right].
\end{equation}

Taking into account equality \eqref{eq15} and the fact that ${e_p} \to 0$, we obtain

\begin{equation}
\Delta {\omega _{sp}} =  - \frac{9}{{20}}e_p^2\frac{{{\epsilon_{\rm{m}}}}}{{{\epsilon^\infty } + 2{\epsilon_{\rm{m}}}}}\omega _{{\rm{sp}}}^{{\rm{sphere}}}.
\end{equation}

\begin{figure}[!t]
\centerline{\includegraphics[width=0.8\textwidth]{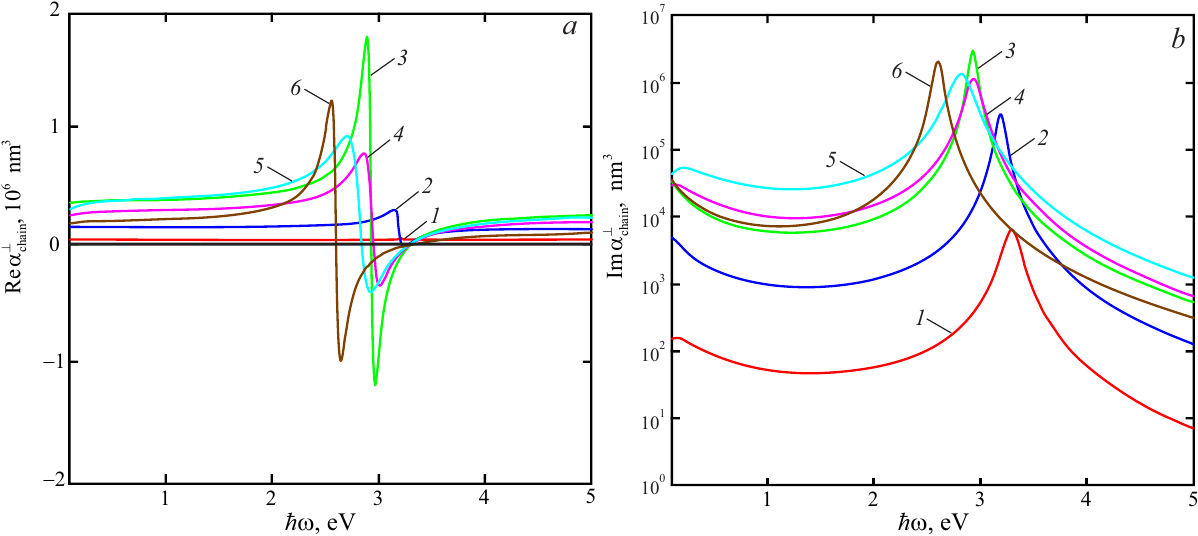}}
\caption{(Colour online) The frequency dependencies for the real (\emph{a}) and imaginary (\emph{b}) parts of the transverse component of the polarizability tensor for the chain of prolate spheroids Au in Teflon under the same values of parameters as in figure~\ref{fig2}.}
\label{fig3}
\end{figure}

The frequency dependencies for the real and imaginary parts of the transverse component of the polarizability tensor of the chain of prolate spheroids are shown in figure~\ref{fig3}.
Let us point out that the maxima of the imaginary part of the chain polarizability experience a ``red'' shift (curves in the sequence $1 \to 2 \to 3$) under an increase in the value of minor semi-axis of spheroids, similarly to the case of an isolated prolate spheroid. At the same time, there is also a red shift under an increase in the value of major semi-axis of spheroid, in contrast to the ``blue'' shift in the case of an isolated spheroid. In addition, all maxima of the imaginary parts of the chain polarizability are shifted to the region of greater frequencies in comparison with similar maxima for the unit spheroid. This phenomenon is similar to the ``blue'' shift of the maxima in the case of two-dimensional (planar) lattices of the nanoparticles \cite{B34}. It should also be pointed out that the chain sums [formulae \eqref{eq9} and \eqref{eq10}] are fast-convegent. Therefore, the calculation of the chain characteristics is not difficult. Moreover, the experiments demonstrate that the values of the sums ${\cal S}_ \bot ^d$ and ${\cal S}_ \bot ^i$ for the cases of long and short chains of prolate spheroids are almost the same. This fact indicates that the proposed theory is applicable to the chains of nanoparticles of different length. The results for the chain of spherical nanoparticles are shown in figure~\ref{fig3},~\emph{a} and \emph{b} (curves 6). By analogy with the case of isolated prolate spheroid, the frequencies of the chain resonance in this case are also less than the frequencies of the chain resonance for the chains of prolate spheroids.

\begin{figure}[!t]
	\centerline{\includegraphics[width=0.8\textwidth]{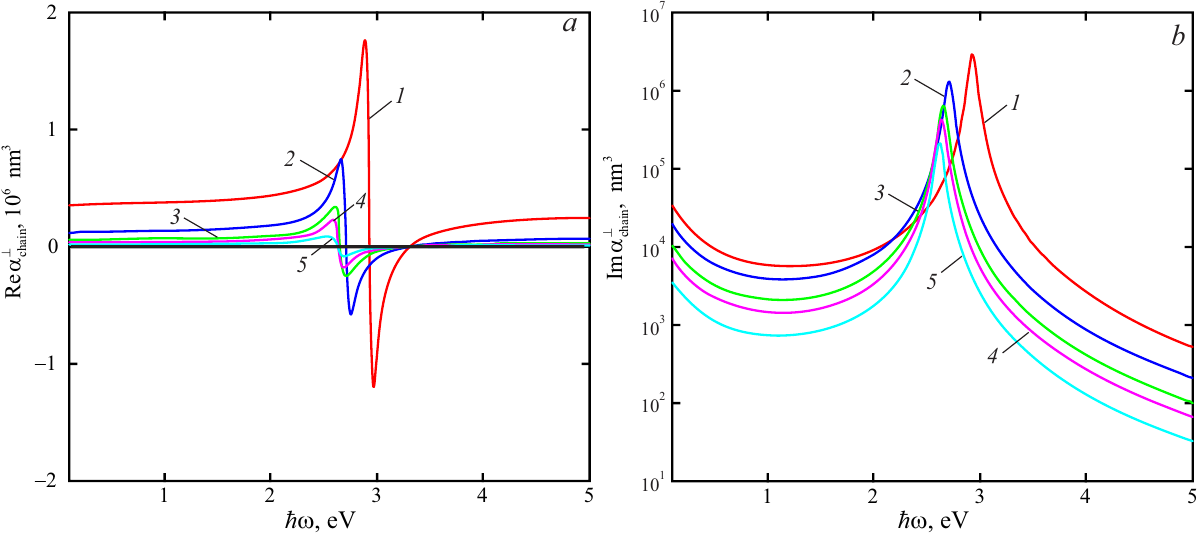}}
	\caption{(Colour online) The frequency dependencies of the real (\emph{a}) and imaginary (\emph{b}) parts of the transverse component of the polarizability tensor for the chain of prolate Au spheroids in Teflon at different distances between particles in the chain ($b = 40\,\,{\rm{nm}}$, $a = 50\,\,{\rm{nm}}$): 1 --- $d = 120\,\,{\rm{nm}}$; 2 --- $d = 150\,\,{\rm{nm}}$; 3 --- $d = 180\,\,{\rm{nm}}$; 4 --- $d = 200\,\,{\rm{nm}}$; 5 --- $d = 240\,\,{\rm{nm}}$.}
	\label{fig4}
\end{figure}
\begin{figure}[!t]
	\centerline{\includegraphics[width=0.80\textwidth]{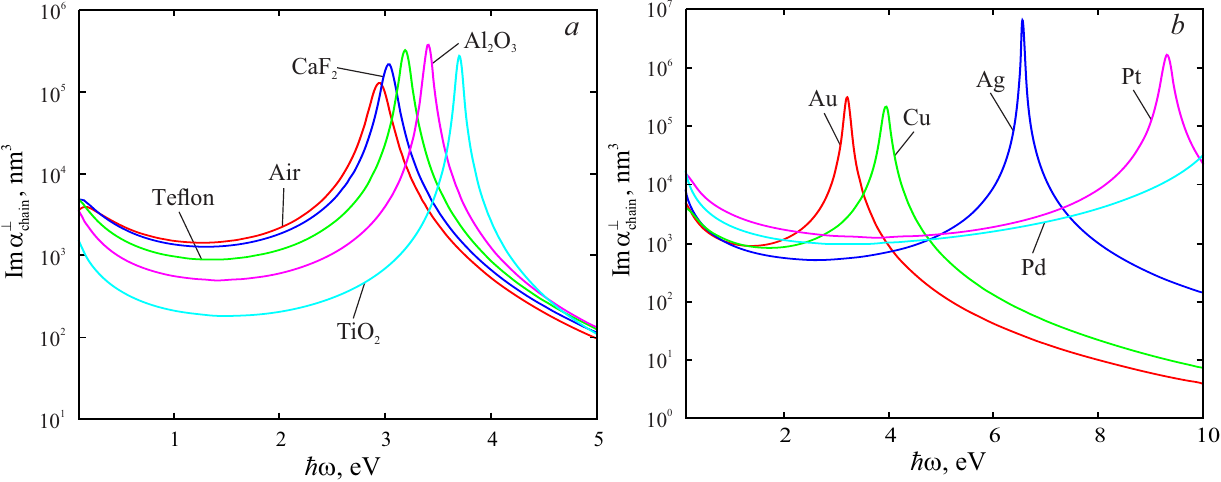}}
	\caption{(Colour online) The frequency dependencies for the imaginary part of the transverse component of the polarizability tensor for the chain of prolate spheroids Au in different matrices (\emph{a}) and the chain of prolate spheroids of different metals in teflon (\emph{b}) under $b = 20\,\,{\rm{nm}}$, $a = 50\,\,{\rm{nm}}$.}
	\label{fig5}
\end{figure}

\begin{figure}[!t]
	\centerline{\includegraphics[width=0.43\textwidth]{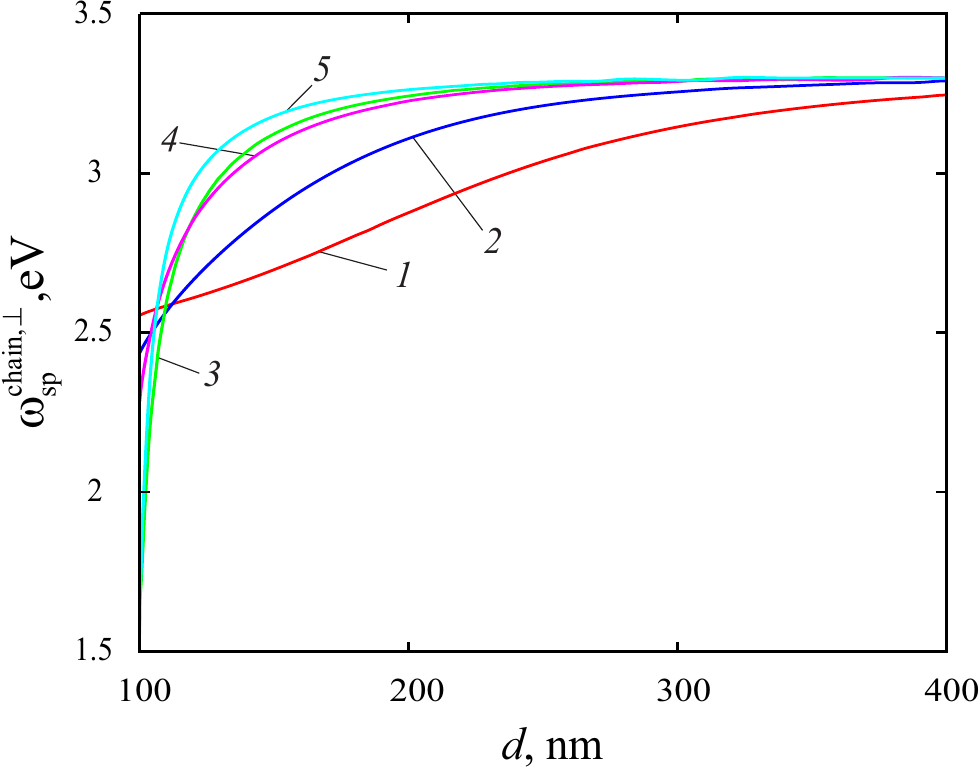}}
	\caption{(Colour online) The dependence between the frequency of the transverse plasmonic resonance in the chain of prolate spheroids Au in Teflon and the period of the chain under the same values of parameters as in figure~\ref{fig2}.}
	\label{fig6}
\end{figure}

The frequency dependencies for the real and imaginary parts of the transverse component of the polarizability tensor for the chain of elongated spheroids in the case of different distances between particles are given in figure~\ref{fig4}. Let us point out that an increase in the distance between particles in the chain results in a decrease in the maximum value of the imaginary part of the chain polarizability and in a ``red'' shift of the maximum itself due to the weakening of the collective interactions.

The frequency dependencies for the imaginary part of the transverse component of the polarizability tensor for the chains of spheroidal particles Au, located in different dielectric media, and the chains, which consist of the nanoparticles of different metals in teflon are given in figure~\ref{fig5}. The calculation results indicate a ``blue'' shift of $\max \left\{ \Im \alpha^\bot _{{\rm{chain}}} \right\}$ under an increase in the permittivity of the surrounding dielectric in the sequence ${\rm{Air}} \to {\rm{Ca}}{{\rm{F}}_2} \to {\rm{Teflon}} \to {\rm{A}}{{\rm{l}}_2}{{\rm{O}}_3} \to {\rm{Ti}}{{\rm{O}}_2}$, moreover, the maxima are located in the near ultraviolet part of the spectrum (figure~\ref{fig4},~\emph{a}) in the case when the environment is oxides ${\rm{A}}{{\rm{l}}_2}{{\rm{O}}_3}$ and ${\rm{Ti}}{{\rm{O}}_2}$. As for the chains of the particles of different metals, it should be pointed out that the maximum of the imaginary part of the polarizability is located in the optical range of the spectrum only in the case of the chains of spheroids Au, and in the ultraviolet part of the spectrum --- in the case of the chains of particles of other metals (Cu, Ag, Pt, Pd) (figure~\ref{fig5},~\emph{b}). This is due to both the differences in the optical characteristics of these metals and the ``blue'' shift of the maxima for the particle chains of the indicated metals compared to isolated particles.

The dependence between the frequency of the transverse resonance in the case of the chains of spheroids of different sizes and the distance between the centers of the particles in the chain is shown in figure~\ref{fig6}. Let us point out that for all considered cases, the frequency of chain resonances grows with increasing distance between the centers of particles in the chain, and under $d \sim 400$~nm, it is practically independent of the sizes (semi-axes of spheroids). Moreover, a sharp drop in the resonant frequency with decreasing magnitude $d$ indicates that the theory proposed in this paper is valid in the case of $\frac{d}{2b} \geqslant 3$ (since $d \sim 100$ nm, and the maximum value $b = 40$~nm).

\section{Conclusions}

The relations have been obtained for the frequency dependence of the transverse component of the polarizability tensor for the chain of metallic nanoparticles on the dielectric substrate which have the shape of prolate spheroids and the size dependence of the frequency of the transverse chain resonance.

It has been shown that the behavior of the maxima of the imaginary parts of the transverse components of the polarizabilities of isolated prolate spheroid and the chain of such particles has both similarities and differences. Thus, with an increasing length of the minor semi-axis in both cases, there is a ``red'' shift of maxima, and with an increase of the major semi-axis, there is a ``blue'' shift for isolated spheroids and the a ``red'' shift for the chain of nanoparticles, and the maxima of the imaginary part of the chain polarizability are shifted to the region of greater frequencies compared to the case of isolated nanoparticles. The indicated differences are a manifestation of the collective effects in the chain of spheroids.

Fast convergence of the chain sums shows the applicability of the theory proposed in the paper to the chains of metallic spheroids of different lengths, and a ``red'' shift of the maxima of the transverse component of the chain polarizability tensor is due to the weakening of the collective interactions under an increase in the distance between particles in the chain (period of the chain).

The results of calculations of the frequency dependence of the imaginary part of the transverse component of the polarizability tensor of the chains of nanoparticle Au in different media indicate a ``blue'' shift of the maxima with increasing dielectric permittivity of the surrounding medium. In the case of the chains of particles of different metals, these maxima are in the ultraviolet region of the spectrum, except for the chains of gold spheroids, which is caused both by the difference in optical properties of metals and the ``blue'' shift of polarizability maxima of the chain of particles, compared to the case of isolated particles.

The limits of applicability of the theory proposed in the paper, determined by the chain period, at which there is a sharp decrease in the frequency of the chain plasmonic resonance, have been established.

\ukrainianpart

\title{Плазмонні резонанси в ланцюжку сфероїдальних металевих наночастинок на діелектричній підкладці}
\author{М.~С.~Манюк\refaddr{label1}, А.~В.~Коротун\refaddr{label1,label2}, В.~І.~Рева\refaddr{label1}, І.~М.~Тітов\refaddr{label1}}
\addresses{
\addr{label1} Національний університет ``Запорізька політехніка'', вул. Університетська, 64,
Запоріжжя, 69063, Україна
\addr{label2} Інститут металофізики ім. Г.~В.~Курдюмова НАН України, бульв. Академіка Вернадського, 36,
  Київ, 03142, Україна
}
%
%
%

\makeukrtitle

\begin{abstract}
\tolerance=3000%
В роботі з використанням наближення локального поля досліджено оптичні
та плазмонні властивості ланцюжків витягнутих металевих сфероїдів на
діелектричній підкладці. Розглянуто випадок, коли сфероїди розташовані так,
що їхня велика вісь лежить у площині підкладки. Отримано співвідношення для
поперечної компоненти тензора ланцюжкової поляризовності та частоти
поперечного ланцюжкового резонансу. Проведено порівняльний аналіз спектральних
зсувів максимумів уявної частини поперечної ланцюжкової поляризовності та
поляризовності ізольованих витягнутих сфероїдів. Вивчено вплив зміни матеріалу
частинок ланцюжка та навколишнього діелектричного середовища на положення
максимумів уявної частини поперечної поляризовності. Встановлено межі
застосування запропонованої в роботі теорії.
\keywords витягнуті металеві сфероїди, ланцюжки наночастинок, діелектрична підкладка, тензор ланцюжкової поляризовності, плазмонний резонанс, ланцюжкові суми

\end{abstract}

\end{document}